\documentclass[12pt]{article}     
\usepackage{graphicx}
\usepackage{amsmath,amsfonts,amsbsy,amssymb,fullpage,makeidx,array,graphicx,epsfig,enumerate}

\title{Editorial note to ``A Homogeneous Universe of Constant Mass and Increasing Radius accounting for the Radial Velocity of Extra--Galactic Nebul\ae'' by Georges Lema\^{\i}tre (1927)} \author{Jean-Pierre Luminet, Laboratoire Univers et Th\'eories\\ Observatoire de Paris - CNRS UMR8102 - Universit\'e Paris Diderot (France)\\email : jean-pierre.luminet@obspm.fr}

\begin{document}
\maketitle

\begin{abstract} This is an editorial note to accompany
printing as a Golden Oldie in the \emph{Journal of General
Relativity and Gravitation}\footnote{See
http://www.mth.uct.ac.za/~cwh/goldies.html.} of the fundamental article 
by Georges Lema\^{\i}tre first published in French in 1927, in which the author provided the first explanation of the observations on the recession velocities of galaxies as a natural consequence of \textit{dynamical} cosmological solutions of Einstein's field equations, and discovered the so--called Hubble law. We analyze in detail the scientific contents of this outstanding work, we describe how it remained unread or poorly  appreciated until 1930, and we list and explain the differences between the 1927 and 1931 versions. Indeed the English translation published in 1931 in MNRAS was not perfectly faithful to the original text -- it was updated. As it turned out very recently, the updates were done by Lema\^{\i}tre himself, but the discrepancies between the two texts caused a temporary stir among historians. Our new translation -- given in the Appendix -- follows the 1927 version exactly. 
\end{abstract}

\bigskip

{\bf \em Introduction}

\bigskip

As already pointed out in a previous Golden Oldie devoted to the Lema\^{\i}tre's
short note of 1931 which can be considered as the true ``Charter'' of the modern
big bang theory [1], although the Belgian scientist was primarily a remarkable
mathematician and a theoretical physicist, he stayed closely related to
astronomy all his life and always felt the absolute need for confronting the
observational data and the general relativity theory. This basic fact explains
why as soon as 1927, while still a beginner in cosmology, he was the first one
to be able to understand the recent observations on the recession
 velocities of galaxies as a natural consequence of \textit{dynamical}
cosmological solutions of Einstein's field equations.\footnote{A number of other
authors such as Hermann Weyl [2], Carl Wirtz [3], Ludwig Silberstein [4], Knut Lundmark
[5] had looked for a relation that fit into the context of de Sitter's static
model which presented \textit{spurious} radial velocities.} Before examining in
detail the contents of his outstanding article, let us summarize the road which, in the
few preceding years, led the young Lema\^{\i}tre to the expanding universe (see
e.g. [6]).

In 1923, the same year as he was ordained as a priest, Georges Lema\^{\i}tre
obtained a 3-year fellowship from the Belgian government, enabling him to study abroad. He spent
 the first year at the University of Cambridge, England, where he studied stellar
astronomy, relativistic cosmology and numerical analysis under the direction of
Arthur Eddington. He spent the second year at Harvard College Observatory in
Cambridge, Massachusets, directed by Harlow Shapley who worked on the problem of
nebul\ae\ . Then he passed to the Massachusetts Institute of Technology
(M.I.T.), where Edwin Hubble and Vesto Slipher were active. The first one
measured the distances of nebul\ae\  by observing variable stars of the Cepheid
type, the second one estimated their radial velocities from their spectral
shifts.

While following closely the experimental work of the American astronomers, who
were going soon to found observational cosmology, Lema\^{\i}tre undertook a PhD
thesis at M.I.T. with his compatriot Paul Heymans as advisor, on the
gravitational field of fluids in general relativity -- a theoretical subject
suggested by Eddington. At the end of 1924, he attended a meeting in Washington
which remained famous since the discovery of Cepheids in spiral nebulas was
announced there by Edwin Hubble; this made it possible to prove the existence of
galaxies external to ours, and Lema\^{\i}tre understood at once that this new
design of ``island universes'' would have drastic consequences for the theories
of relativistic cosmology.

On July 1925, his American stay ended and Lema\^{\i}tre had to go back to
Belgium. In this same decisive year for observational cosmology, Lema\^{\i}tre
obtained his first notable scientific results, concerning the cosmological
solution found by De Sitter [7]. In the first article [8] he demonstrated how he could introduce new coordinates for the
De Sitter universe which made the metric no more static, with a space of zero
curvature and a scale factor depending exponentially on time. This metric would
be used twenty years later by the keenest adversaries of the theory of the
expanding universe in the framework of ``steady-state'' models [9], and still
later in the 1980's to describe the hypothetical inflationary phase of the very
early universe, see e.g. [10]. In the second article [11] he deduced that the relation between the relative speed of
test-particles and their mutual distances in the De Sitter universe was linear.
It was the first time that the cosmological constant (when it is positive) was
seen allotting the role of a ``cosmic repulsion'' forcing the worldlines of
particles to recede with time. However, although he found this non--static feature to be promising because
of its connection to the redshifts of nebul\ae\ , he also realized that the
model resulted in an infinite Euclidean space, that he considered inadmissible:
as a neo-Thomist he did not accept the actual infinite and remained faithful to
the finitude of space and matter throughout his career. Thus he had to seek for
an alternative explanation, involving a truly non--static
and spatially closed solution of Einstein's equations.

In 1926-27, Lema\^{\i}tre went again to the United States, where he remained at
M.I.T. during three quarters of the academic year. Back in Europe in June 1927, he was informed by
letter that he got his PhD [12], having been exempted of oral defense. The same
year, he was appointed professor at the University of Louvain and published his
great article on the expanding universe.

\bigskip

{\bf \em Recession of galaxies and expanding universe}

\bigskip

Since 1912, Vesto Slipher had undertaken a program of measurement of the radial
velocities of spiral nebulae. Interpreted in terms of the Doppler effect, the
shifts in frequency (or wavelength) implied a radial speed of displacement of
the source compared to the observer. Radial speeds were thus indirectly measured
by spectroscopy. By 1917, Slipher (see [13] and references therein) had analyzed
the spectra of 25 spiral nebul\ae, which he had observed at Lowell Observatory
in Flagstaff, Arizona; 21 of them presented redshifts that could be
interpreted as a systematic motion of recession (the exceptions were M81 and 3
galaxies from the Local Group). However, nobody suspected yet the repercussions
that these preliminary data would have soon for the whole of cosmology, mainly
due to the fact that the debate on whether spiral nebul\ae\ were island
universes went on. The evidence for the redshifts mounted mainly due to
Slipher's efforts, and by 1923 reached a score of 36 among 41 spiral nebul\ae\ .

Slipher never published his final list,\footnote{For details see [14].} but it
was given in Arthur Eddington's book of 1923 [15], who noticed that ``one of the
most perplexing problems in cosmogony is the great speed of spiral nebul\ae\ .
Their radial velocities average about 600 km. per sec. and there is a great
preponderance of velocities of recession from the solar system''. The
influential British astronomer suggested that effects due to the curvature of space-time should be looked for
and referred to De Sitter's model for a possible explanation.

Thanks to his various stays at Cambridge, England, and at M.I.T. (where he met
Slipher personally), Lema\^{\i}tre was perfectly informed of these preliminary results, and he wanted
to take account of the available data by using a \textit{new} cosmological
solution of Einstein's equations.

As the title of his 1927 article clearly states, Lema\^{\i}tre was
able to connect the expansion of space arising
naturally from the non--static cosmological solutions of general relativity with
the observations of the recession velocities of extragalactic nebul\ae.

He begins to review the dilemma between the De Sitter and Einstein
universe models. The De Sitter model ignored the existence of matter;
however, it emphasized the recession velocities of spiral nebul\ae\  as a simple consequence
of the gravitational field. Einstein's solution allowed for the presence of
matter and led to a relation between matter density and the radius of the space
-- assumed to be a positively curved hypersphere; being strictly static due an
adjustment of the cosmological constant, it could not, however, explain the
recession of the galaxies. Lema\^{\i}tre thus looks for a new solution of the
relativistic equations combining the advantages of the Einstein and De Sitter
models without their inconveniences, i.e. having a material content and
explaining at the same time the recession velocities.

For this, in the next section he assumes a positively curved space (as
made precise in a footnote, with ``elliptic topology'', namely that of the projective space
$\bf{P}^3$ obtained by identification of antipodal points of the
simply-connected hypersphere $\bf{S}^3$; see [16] for an explanation of such a
choice) with the radius of curvature $R$ (and consequently the
matter density $\rho$) being a function of time $t$, and a non-zero cosmological constant $\lambda$. From
Einstein's field equations he obtains differential equations (eq. (2)-(3)) for
$R(t)$ and $\rho(t)$ almost identical to those previously obtained by Friedmann
[17] (at the time Lema\^{\i}tre was not aware of Friedmann's work, see below).
The difference is that Lema\^{\i}tre supposes the conservation of energy (eq.
(4)) -- this is the first introduction of thermodynamics in relativistic
cosmology -- and he includes the pressure of radiation as well as the term of
matter density into the stress-energy tensor (he rightly considers the matter
pressure to be negligible). Lema\^{\i}tre emphasizes the importance of radiation
pressure in the first stages of the cosmic expansion. Now it is well known that,
within the framework of big bang models, the approximation of zero pressure is
valid only for times posterior to the big bang for approximately four hundred
thousand years. Just like Einstein and De Sitter, Friedmann had made the
assumption that the term of pressure in the stress-energy tensor was always
zero. The equations derived by Lema\^{\i}tre are thus more general and
realistic.

Lema\^{\i}tre  shows how the Einstein and De Sitter
models are particular solutions of the general equations. Next he chooses as
initial conditions  $R' = R'' = 0, R = R_0$ at $t = - \infty$ and he adjusts the
value of the cosmological constant such that $\lambda = 1/R_0$, in the same way
Einstein had adjusted the value of $\lambda$ in his static model with constant
radius.

As a consequence, the exact solution he obtains in eq. (30) describes a
monotonous expanding universe, which, when one indefinitely goes
back in time, approaches in an asymptotic way the Einstein
static solution, while in the future it approaches asymptotically an
exponentially expanding De Sitter universe.

This model, deprived of initial singularity and, consequently, not possessing
a definite age -- as well as the ``monotonous solution of second species'' found
earlier by Friedmann -- will be later baptized
the Eddington-Lema\^{\i}tre's model (see below).

Lema\^{\i}tre does not provide a graph for $R(t)$ but gives numerical values in
a table going from $t = - \infty$ to $+ \infty$. For the sake of clarity our
figure 1 depicts such a graph.

 \begin{figure}[h!]
 \begin{center}
\includegraphics[scale=0.5]{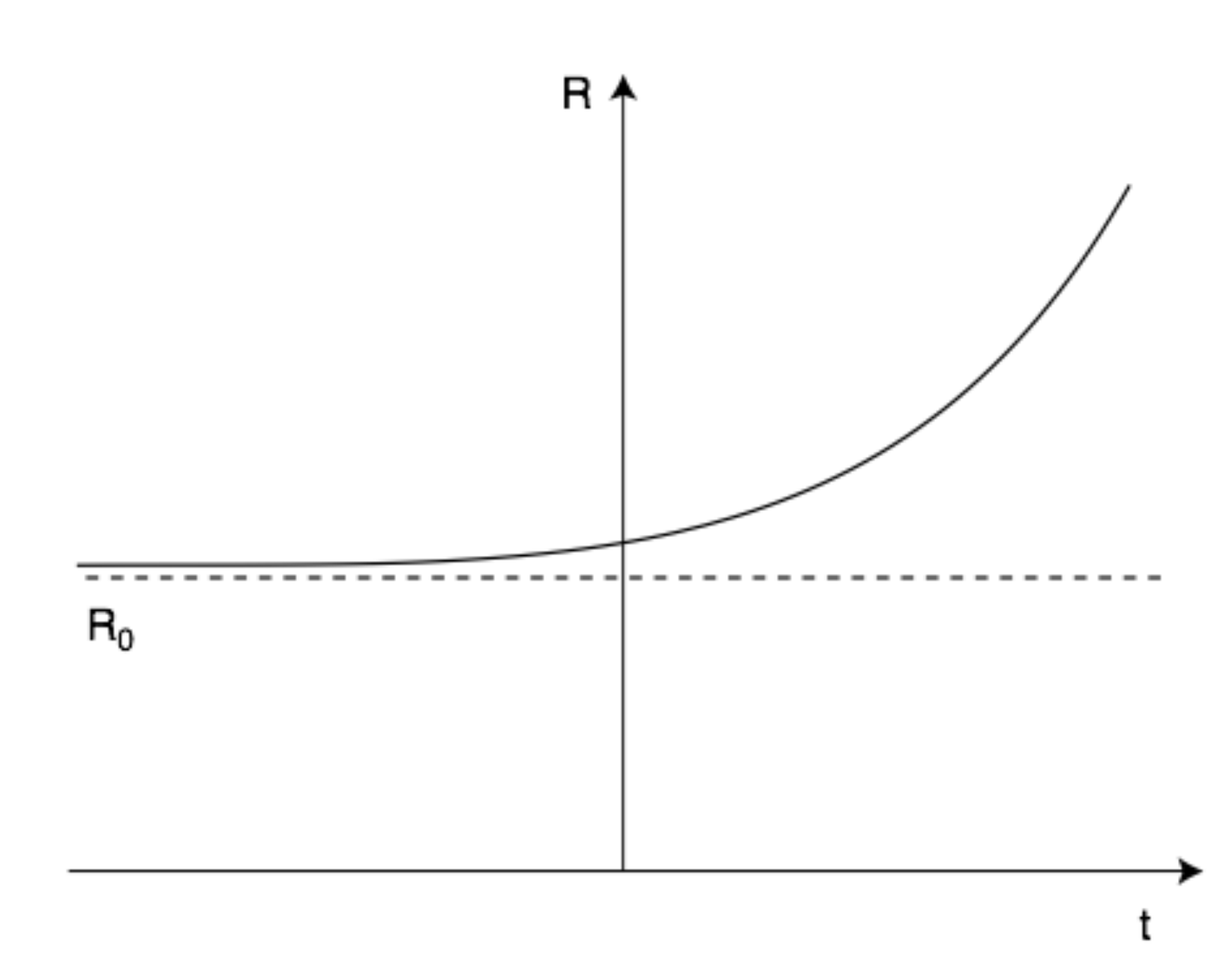}
 \caption{
The 1927 Lema\^{\i}tre's universe model, later
named Eddington--Lema\^{\i}tre. The radius $R_0$ of the static Einstein
hypersphere is reached asymptotically for $t=- \infty$. The origin of cosmic
time is arbitrary, thus the model does not pose any problem of age.
 }
 \end{center}
 \end{figure}

Lema\^{\i}tre conceived the static Einstein universe as a kind of pre-universe
out of which the expansion had grown as a result of an instability. As a
physical cause for the expansion he suggested the radiation pressure itself, due to its infinite accumulation in a closed static universe, but he did not develop this (erroneous) idea.

While giving preference to this particular model in his article, Lema\^{\i}tre nevertheless
calculated separately the whole of dynamical homogeneous cosmological solutions, since he had the
general formula (eq. (11)) making it possible to calculate the time evolution of
all the homogeneous isotropic models with positive curvature. The Lema\^{\i}tre
archives at the University of Louvain keep a red pad with the inscription
``1927'', which contains the galley proofs of his article, some notes in
handwriting connected with the paper, and two diagrams which (unfortunately) do
not appear in any of his publications. These diagrams depict the time evolution
of the space scale factor depending on the value of the cosmological constant
for all homogeneous and isotropic solutions of Einstein's equations with
positive curvature of space.

 \begin{figure}[h!]
 \begin{center}
 \includegraphics[scale=0.5]{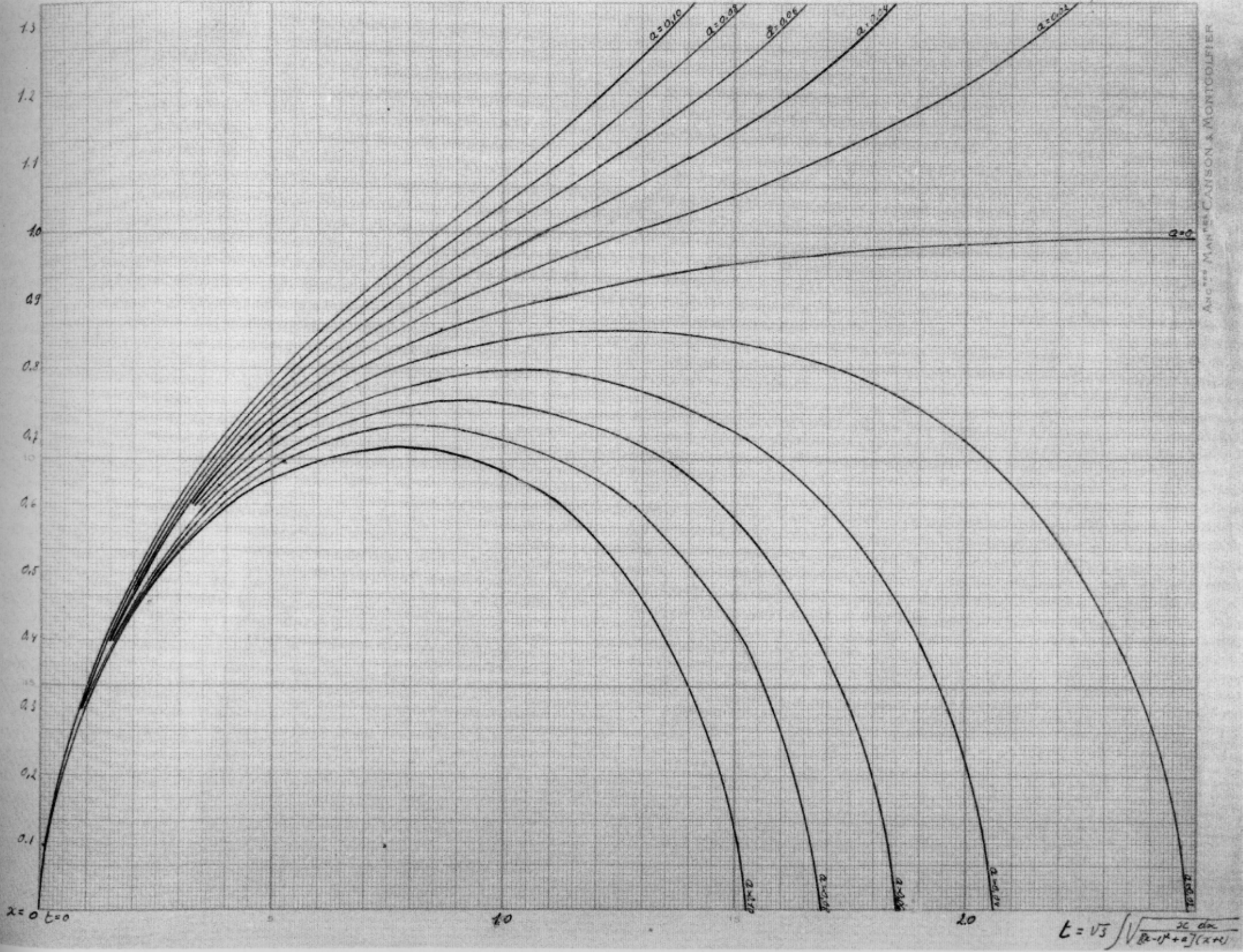}
 \caption{
Handwritten graph by Lema\^{\i}tre. This extraordinary diagram, plotted by
Lema\^{\i}tre in 1927, but unpublished until 1998 [20],
depicts the time evolution of the radius of the universe
with the cosmological constant (denoted $a$), for a space of positive curvature. All the models start with a singularity
in $(x = 0, t = 0)$. For a sufficiently large cosmological constant $a$, the
universe becomes open. The most recent cosmological data are compatible with a
Lema\^{\i}tre's solution with positive curvature and accelerated expansion (top
curve).
 }
 \end{center}
 \end{figure}

As mentioned above, the 1927 article does not refer to the work of Friedmann,
published in \textit{Zeitschrift f\"{u}r Physik} -- although one of the
best known journals in theoretical physics at that time. This absence seems strange
if one remembers the two notes by Einstein published in the same review [18],
which had been largely discussed in the scientific community. A plausible
explanation is that Lema\^{\i}tre could not read the German [19].
Friedmann's articles were pointed out to Lema\^{\i}tre by Einstein himself,
during their meeting at the 1927 Solvay Conference. The reference to Friedmann
thus appears for the first ime in a text of 1929 written in French, \textit{La
grandeur de l'espace} [21], in which Lema\^{\i}tre thanks ``Mr. Einstein for the
kindness that he showed by announcing
to me the important work of Friedmann which includes several of the results
contained in my note on a homogeneous universe''. The reference will also appear
in the 1931 English translation of Lema\^{\i}tre's article, see below.

The exceptional interest of Lema\^{\i}tre's work is to provide the first
interpretation of cosmological redshifts as a natural effect of the expansion of
the universe within the framework of general relativity, instead of a real
motion of galaxies: as it is written down in eq.(23), space is constantly
expanding and consequently increases the apparent separations between galaxies.
This idea will prove to be one of the most profound discoveries of our time.

The relation of proportionality (23) between the recession velocity and the
distance is an approximation valid at not too large distances which can be used
``within the limits of the visible spectrum''.  Then, using the available
astronomical data, Lema\^{\i}tre provides the explicit relation of
proportionality in eq. (24), with a factor 625 or 575 km/s/Mpc, depending on his
choice of observations which presented an enormous scatter.  This is the first
determination of the so-called Hubble law and the Hubble constant, that should
as well have been named Lema\^{\i}tre's law.

For this the Belgian scientist uses a list of 42 radial velocities compiled by
Gustav Str\"{o}mberg, a Swedish astronomer at the Mount Wilson
Observatory,\footnote{Str\"{o}mberg [22] relied himself on redshifts measured by
Slipher and included some globular clusters in addition to spiral nebul\ae\ .}
and deduces their distance from a recent empirical formula between the distance
and the absolute magnitude provided by Hubble [23], who himself took them from
Hopmann [24]. Eventually, Lema\^{\i}tre is able to give the numerical figures
for the initial and present-day values of the radius of the universe, resp. $R_0
= 2,7 \times 10^8 pc$ and $R = 6 \times 10^9 pc$. At the very end he points out
that the largest part of the universe will be forever out of reach of the
visible spectrum, since the maximum distance reached by the Mt Wilson telescope
is only $R/120$, whereas for a distance only greater than $R/11,5$ the whole
visible spectrum is displaced into the infrared -- he could not imagine the
space era with infra-red and submillimeter telescopes placed on board of
satellites.

We have seen above that Lema\^{\i}tre knew already all the solutions of
Einstein's equations for homogeneous and isotropic universes. The reason why he
privileged a very particular model, adjusting the cosmological constant in order
to have no beginning of time, is due to his overestimate of the Hubble constant:
as is well known, the latter gives an order of magnitude of the duration of the
expansion phase; with the estimate of about 600 km/s/Mpc found by Lema\^{\i}tre,
this period is about one billion years only, a number less than the age of the
Earth estimated by the geologists of the time. Thus the model with exponential
expansion and no beginning allowed to reconcile the theory with both
astronomical and geological data.

\bigskip

{\bf \em First reactions}

\bigskip

The significance of Lema\^{\i}tre's work has remained mostly unnoticed for three
years, not exclusively (but partly) due to the fact that it was published in French in an ``obscure and
completely inaccessible journal'', as is
sometimes claimed [25], instead of one of the prestigious astronomical journals
of the time.\footnote{The paper was reprinted later in 1927 in vol. 4 of
\textit{Publications du Laboratoire d'Astronomie et de G\'eod\'esie de
l'Universit\'e de Louvain}, still less suited for widespread dissemination.} As
rightly pointed out by Lambert [26], the \textit{Annales de la Soci\'et\'e
Scientifique de Bruxelles} published some articles in English, had an excellent
scientific level and therefore were displayed in a large number of academic
libraries and observatories all around the world; also French could be read by a
much larger scientific audience than today. Indeed, the main obstacle to a
larger diffusion of Lema\^{\i}tre's article was that most of the physicists of
the time, such as Einstein and Hubble, could not accept
the idea of a non--static universe. This was not the case
with Eddington; unfortunately, his former mentor, to whom Lema\^{\i}tre had sent a
copy, either forgot to read it in time, or he had not understood its importance.

{}From 24 to 29 October 1927
the Fifth Solvay Conference in Physics took place in Brussels,
one of the great meetings of world science. The Solvay Conference was devoted to
the new discipline of quantum mechanics, whose problems disturbed many
physicists. Among them was  Einstein. For Lema\^{\i}tre, it was the opportunity
to discuss with the father of general relativity. He later reported himself on
this meeting: ``While walking in the alleys of the Parc L\'eopold, [Einstein]
spoke to me about an article, little noticed, which I had written the previous
year on the expansion of the universe and which a friend had made him read.
After some favorable technical remarks, he concluded by saying that from the
physical point of view that appeared completely abominable to him. As I sought
to prolong the conversation, Auguste Piccard, who accompanied him, invited me to
go up by taxi with Einstein, who was to visit his laboratory at the University
of Brussels. In the taxi, I spoke about the speeds of nebulas and I had the impression that Einstein was hardly aware of the astronomical facts. At the university, everyone began to speak in German'' [27].
Einstein's response to Lema\^{\i}tre shows the same unwillingness to change his
position that characterized his former response to Friedmann (see e.g. [28]): he
accepted the mathematics, but not a physically expanding universe!

In 1928 H. P. Robertson published an article [29] in which he wanted to replace
De Sitter's metric by a  ``mathematically equivalent in which many of the
apparent paradoxes inherent in [De Sitter's solution] were eliminated''. He got
the formula $v = cd/R$ where $d$ is the distance of the nebula and $R$ the
radius of curvature of the universe, but in the framework of a \textit{static}
solution. Robertson used the same set of observations as had been taken by
Lema\^{\i}tre\footnote{He did not know the Lema\^{\i}tre's articles of 1925 and
1927.} and that would be taken by Hubble one year later. From this he calculated
$R = 2 \times 10^{27}$ cm, and a proportionality constant of 464 km/s/Mpc (that
he did not calculate, the figure can be found in [30]). The main interest of
Robertson's work (see also [31]) is that he was the first to search in detail
for all the mathematical models satisfying a spatially homogeneous and isotropic
universe -- which imply strong symmetries in the solutions of Einstein's
equations.

In 1929, Hubble [32] used the experimental data on the Doppler redshifts mostly
given by Slipher (who was not quoted) and found a linear velocity-distance
relation $v = Hr$ with $H = 465 \pm 50$ km/s/Mpc for 24 objects and $513 \pm 60$
km/s/Mpc for 9 groups. The law was strictly identical to Lema\^{\i}tre's
Eq.(24), with almost the same proportionality factor, but Hubble did not make
the link with expanding universe models. He stated ``The outstanding feature,
however, is the possibility that the velocity-distance relation may represent
the De Sitter effect''.  In fact Hubble never read Lema\^{\i}tre's paper; he
interpreted the galaxy redshifts as a pure Doppler effect (due to a proper
radial velocity of galaxies) instead of as an effect of space expansion. And
throughout his life he would stay skeptical about the general relativistic interpretation of his
observations. For instance, in the 202 pages of his book of 1936 \textit{The
Realm of the Nebulae} [33], he tackled the theoretical interpretation of the
observations only in a short ultimate paragraph on
page 198, in which he quoted Einstein, De Sitter, Friedmann, Robertson, Tolman
and Milne. As pointed out by his biographer G. Christianson, Hubble was chary of
``all theories of cosmic expansion long after most astronomers and physicists
had been won over. When queried about the matter as late as 1937, he sounded
like an incredulous schoolboy: `Well, perhaps the nebulae are all receding in
this peculiar manner. But the notion is rather startling{' }'' [34]. Indeed the
fact that the expansion of the universe was discovered by Hubble is a myth that
was first propagated by his collaborator Humason as soon as 1931 (see e.g. [35])
and Hubble himself, who was fiercely territorial; in a letter to De Sitter dated
21 August 1930, he wrote ``I consider the velocity-distance relation, its
formulation, testing and confirmation, as a Mount Wilson contribution and I am
deeply concerned in its recognition as such'' (quoted in [36]).

One month only after Hubble's article, Tolman joined the game of searching for
an explanation of recession velocities, but still in the framework of a static
solution [37], as he said ``the correlation between distance and apparent radial
velocity of  the extra--galactic nebulae obtained by Hubble, and the recent
measurement of the Doppler effect for a very distant nebula made by Humason at
the Mount Wilson Observatory, make it desirable to consider once more the
theoretical relations between distance and Doppler effect which could be
expected from the form of line element for the universe proposed by De Sitter''.
One year later, Tolman published another article [38] where he suggested that
the expansion was due to the conversion of matter into radiation, an idea
already proposed by Lema\^{\i}tre in his 1927 article, who again was not quoted.

A new opportunity for the recognition of Lema\^{\i}tre's model arose early in
1930. In January, in London, a discussion between Eddington and De Sitter took
place at a meeting of the Royal Astronomical Society. They did not know how to
interpret the data on the recession velocities of galaxies. Eddington suggested
that the problem could be due to the fact that only static models of the
universe were hitherto considered; he nicely formulated the situation as follows: ``Shall we put a little motion into
Einstein's world of inert matter, or shall we put a little matter into de
Sitter's Primum Mobile?'' [39], and called for new searches in order to explain
the recession velocities in terms of dynamical space models.

Having read the report of the meeting of London, Lema\^{\i}tre understood that
Eddington and De Sitter posed a problem which he had solved three years earlier.
He thus wrote to Eddington to remind him about his communication of 1927 and
requested him to transmit a copy to de Sitter: ``Dear Professor Eddington, I
have just read the February n$^\circ$ of the \textit{Observatory} and your
suggestion of investigating non statical intermediary solutions between those of
Einstein and De Sitter. I made these investigations two years ago. I consider a
universe of curvature constant in space but increasing with time. And I
emphasize the existence of solution in which the motion of the nebul\ae\  is
always a receding one from time minus infinity to plus infinity.''\footnote{From a copy
kept at the Archives Lema\^{\i}tre of Louvain-la-Neuve, quoted in [26].}
Lema\^{\i}tre precised: ``I had occasion to speak of the matter with Einstein
two years ago. He told me that the theory was right and is all which needs to be
done, that it was not new but had been considered by Friedmann, he made critics against which
he was obliged to withdraw, but that from the physical point of view it was
`tout \`a fait abominable' '' (quoted in [40]).

The British astrophysicist was one of the most prominent figures of science at
the time, and was in the best possible position to play a key role in the
recognition of the Lema\^{\i}tre's results. This time he paid attention to
Lema\^{\i}tre's contribution, dispatched a copy to De Sitter in Holland and H.
Shapley in the United States. Eddington was somewhat embarrassed. According to
George McVittie, at the time a research student of Eddington working with him on
the stability of the Einstein's static model, ``[I remember] the day when
Eddington, rather shamefacedly, showed me a letter from Lema\^{\i}tre which
reminded Eddington of the solution to the problem which Lema\^{\i}tre had
already given. Eddington confessed that although he had seen Lema\^{\i}tre's
paper in 1927 he had forgotten completely about it until that moment'' (quoted
in [40]).

On March 19th, Eddington accompanied his invoice of Lema\^{\i}tre's paper to De
Sitter in Leiden by the following comment: ``It was the report of your remarks
and mine at the [Royal Astronomical Society] which caused Lema\^{\i}tre to write
to me about it. At this time, one of my research students, McVittie, and I had
been worrying at the problem and made considerable progress; so it was a blow to
us to find it done much more completely by Lema\^{\i}tre (a blow attenuated, as
far as I am concerned, by the fact that Lema\^{\i}tre was a student of mine)''
(reported in [41]).

De Sitter answered Lema\^{\i}tre very favorably
in a letter dated March 25th, 1930, and the Belgian physicist replied to him on
April 5th (these letters are fully displayed in [42]). In late May, De Sitter
published a discussion about the expansion of the universe [43], where he wrote
``A dynamical solution of the equations (4) with the line-element (5) (7) and
the material energy tensor (6) is given by Dr. G. Lema\^{\i}tre in a paper
published in 1927, which had unfortunately escaped my notice until my attention
was called to it by Professor Eddington a few weeks ago.''

On his side, Eddington reworked his communication to the following meeting of
the Royal Astronomical Society in May, to bring Lema\^{\i}tre's work to the
attention of the world [44].  Then he published an important article [45] in
which he reexamined the Einstein static model and discovered that, like a pen
balanced on its point, it was unstable: any slight disturbance in the
equilibrium would start the increase of the radius of the hypersphere; thus he
adopted Lema\^{\i}tre's model of the expanding universe -- which will be henceforward
referred to as the Eddington--Lema\^{\i}tre model -- and calculated that the
original size of the Einstein universe was about 1200 million light years, of
the same order of magnitude as that estimated by Lema\^{\i}tre in 1927.
Interestingly enough, Eddington also considered the possibility of an initial
universe with a mass $M$ greater or smaller than the mass $M_E$ of the Einstein
model, but he rejected the two solutions, arguing that, for $M > M_E$, ``it
seems to require a sudden and peculiar beginning of things'', whereas for $M <
M_E$, ``the date of the beginning of the universe is uncomfortably recent''.

Eventually, Eddington sponsored the English translation of the 1927
Lema\^{\i}tre's article for publication in the \textit{Monthly Notices of the
Royal Astronomical Society} [46].

Then, with the support of Eddington and De Sitter, Lema\^{\i}tre suddenly rose
to become a celebrated innovator of science. He was invited to London in order
to take part in a meeting of the British Association on the relation between the
physical universe and spirituality. But in the meantime
he had considerably progressed in his investigations of relativistic
cosmologies, and instead of promoting his model of 1927, he dared to propose
that the Universe expanded from an initial point which he called the ``Primeval
Atom''. Then cosmology experienced a second paradigmatic shift [47].

\bigskip

{\bf \em The English translation and discrepancies}

\bigskip

A great deal has been written on the topic of who really discovered the
expanding universe [48]. The French astronomer Paul Couderc [49] was probably
the first one to rightly underline the priority of Lema\^{\i}tre over Hubble,
but since Lema\^{\i}tre himself never claimed any priority (see
[50] for more details), the case was not much discussed.

An intriguing discrepancy between the original French article and its English
translation had already been quoted by various authors (e.g. [40--42]):  the
important paragraph discussing the observational data and eq. (24) where
Lema\^{\i}tre gave the relation of proportionality between the recession
velocity and the distance (in which the determination of the constant that later
became known as Hubble's constant appears) was replaced by a single sentence:
``From a discussion of available data, we adopt $R'/R = 0, 68 \times 10^{-27}
cm^{-1}$''. It was found curious that the crucial paragraphs assessing the
Hubble law were dropped so that,  either due to Eddington's
blunder\footnote{Until very recently the identity of the translator was not
assessed, generally assumed to be that of Eddington himself.} or some other
mysterious reason, Lema\^{\i}tre was never recognized as the discoverer of the
expansion of the universe. \textit{De facto} Lema\^{\i}tre was eclipsed and
multitudes of textbooks proclaim Hubble as the discoverer of the expanding
universe, although Hubble himself never believed in such an explanation [51].

Suddenly, in 2011, a burst of accusations has flared up against Hubble, from the
suspicion that a censorship was exerted either on Lema\^{\i}tre by the editor of
the M.N.R.A.S. [52] or on the editor by Hubble himself [36] -- suspicion based
on the ``complex personality'' of Hubble, who strongly desired to be credited
with determining the Hubble constant.

The  controversy was ended by Mario Livio, from the \textit{Space Telescope
Institute} [53], with the help of the Archives Lema\^{\i}tre at Louvain and the
Archives of the Royal Astronomical Society (see also [26] for additional
details). It is not the scope of  the present note to enter into the
explanations that solve the conundrum, it is sufficient to say that it is now
certain that Lema\^{\i}tre himself translated his article, and that he chose to
delete several paragraphs and notes without any external pressure. On the
contrary, he was encouraged to \textit{add} comments on the subject; but the
Belgian scientist, who had indeed new ideas, preferred to publish them in a
separate article, published in the same issue of M.N.R.A.S. [54].

For the present purpose it is much more interesting to list in detail
all the discrepancies -- as far as we know a little work that has never been
done -- in order to better understand how the preoccupations of Lema\^{\i}tre
had changed since 1927, and how the question which he had in mind in 1931 was
less the expansion of space than the deep cause of it, how it started and how
the first galaxies could form.

\begin{itemize}

\item Section 1, first paragraph

The footnote ``We consider simply connected elliptic space, i.e. without
antipodes'' is suppressed from the 1931 translation.

As soon as 1917, De Sitter  [17][55] distinguished the spherical space $\bf{S}^3$
and the projective space $\bf{P}^3$, that he called the elliptical space. As
recalled by Lema\^{\i}tre in the first paragraph, $\bf{P}^3$ has a (comoving)
volume $\pi ^2~R^3$ instead of $2 \pi ^2~R_0^3$ for $\bf{S}^3$, and the longest
closed straight line is  $\pi R$ instead of  $ 2\pi R$. The main cosmological
difference is due to the presence, in $\bf{S}^3$, of an antipodal point
associated to any point, and in particular to the observer, at a distance of  $
\pi R$ precisely. This was considered as an undesirable fact, so that
cosmological models with $\bf{P}^3$ seemed preferable than those with
$\bf{S}^3$.

Eddington [15] also referred to elliptical space as an alternative  more
attractive than $\bf{S}^3$, and Lema\^{\i}tre also adopted this point of
view.\footnote{Note that elliptical space is not simply connected but multiply
connected, see e.g. [16].} We can infer that he suppressed his footnote because
in any case, topology has no influence on the dynamics, which was the very
purpose of his work, and because in the meantime
he had published an extended discussion on the subject [54], which he merely
points out in reference 4 of the 1931 version.

\item Section 1, second paragraph

In the 1931 translation, the original sentence ``[\ldots] it is of great
interest as explaining the fact that extragalactic nebul\ae\ \  seem to recede
from us with a huge velocity [\ldots]'' is replaced by ``[\ldots] it is of
extreme interest as explaining quite naturally the observed receding velocities
of extragalactic nebul\ae\ \ [\ldots]'' to acknowledge the fact  that, due to the post--1927 observational work of Hubble and Humason, the receding
velocities had acquired a firm observational status.

\item Section 1, third paragraph

The sentence ``This relation forecasted the existence of masses enormously
greater than any known when the theory was for the first time compared with the
facts'' is replaced by ``This relation forecasted the existence of masses
enormously greater than any known at the time''.

\item Section 1, third paragraph

The footnote giving reference to Hubble's article of 1926 is suppressed because it is no more up-to-date.

\item Section 1, sixth paragraph

The two footnotes are suppressed. They both give geometrical details and
subtleties about the De Sitter solution that Lema\^{\i}tre probably
judged not appropriate for a journal such as \textit{M.N.R.A.S}, more devoted to astronomy than to
geometry. These details came mainly from an article by K. Lanczos and the 1925
articles of Lema\^{\i}tre himself. In the 1931 version, he added at the end of
the article the bibliographic references to Lanczos and himself without
development, and added references to H. Weyl and P. du Val.

\item Section 2

Between eqs. (4) and (5) the sentence ``It is suitable for an interesting
interpretation'' has disappeared for the sake of economy.

\item Section 4

The paragraphs from ``Radial velocities of 43 extra-galactic nebul\ae\ \
[\ldots]''  up to ``This relation enables us to calculate $R_0$'', as well as
the three footnotes, are suppressed and replaced by ``From a discussion of
available data, we adopt $R'/R = 0, 68 \times 10^{-27} cm^{-1}$''. This is
precisely the part of the 1927 article where Lema\^{\i}tre discusses the
astronomical data on the redshifts, the errors in the distance estimates,  where
he gives the relation of proportionality between the velocity and distance, and
in footnotes, the references to Str\"{o}mberg and Lundmark, as well as his
calculation of two possible values of the constant of proportionality of 575 and
670, depending on how the data are grouped. The original eq. (24) is truncated
to a pure numerical one,  whereas the original gives precisely what is called
the Hubble's law.

In a letter dated 9 March 1931 addressed to William H. Smart, the editor of
\textit{M.N.R.A.S.},  Lema\^{\i}tre writes: ``I send you a translation of the
paper. I did not find advisable to reprint the provisional discussion of radial
velocities which is clearly of no actual interest, and also the geometrical one,
which could be replaced by a small bibliography of ancient and new papers
on the subject'' (quoted in [53]). The choice of  Lema\^{\i}tre is quite
comprehensible because the data he used in 1927 gave only very imperfectly the
linear relation $v = Hd$, whereas in 1931 the new data from Hubble allowed to
validate this relationship in a much more precise manner, see figure 3 for comparative
plots. Also because, as he explained himself in 1950, in 1927 he had not at his
disposal data concerning clusters of galaxies, and he added that Hubble's law
could not  be proved without the knowledge of the clusters of galaxies'' [56].
Here we find again one of the characteristic features of Lema\^{\i}tre's
personality already mentioned, namely the crucial importance he always gave to
experimental data.

 \begin{figure}[h!]
 \begin{center}
 \includegraphics[scale=0.3]{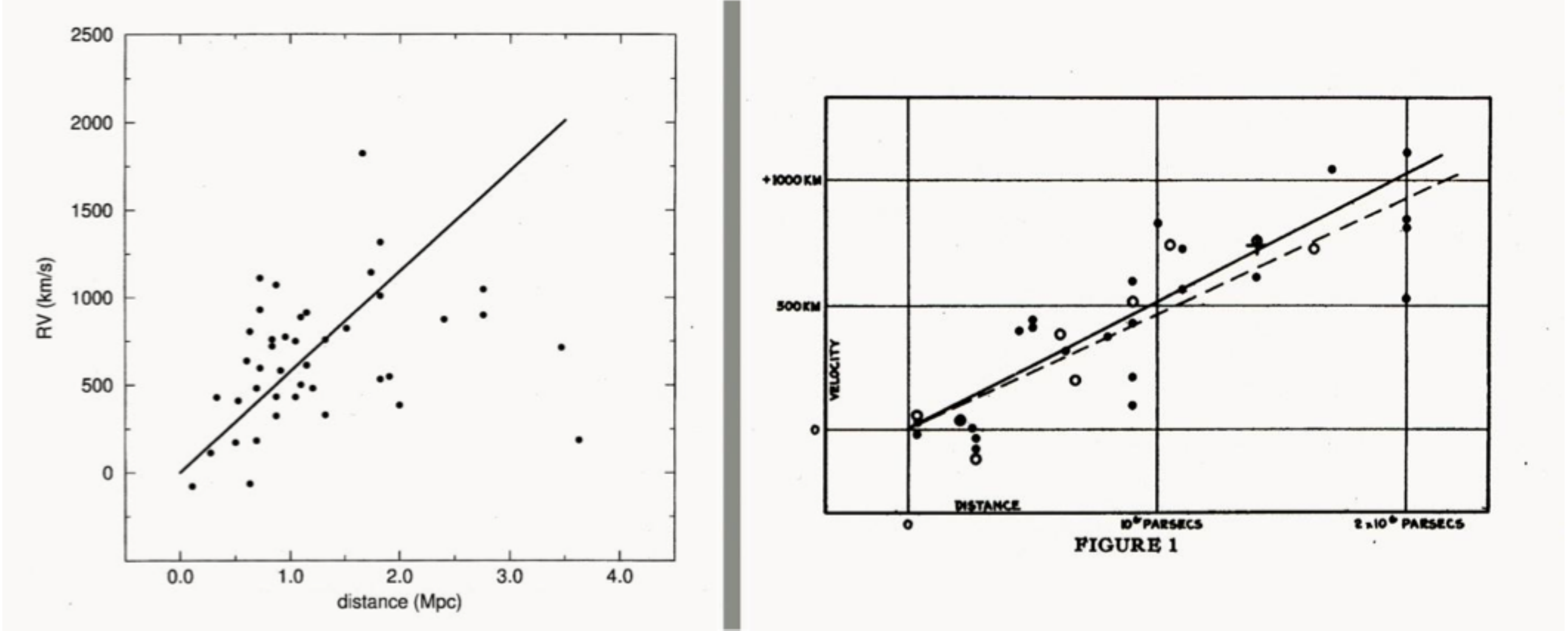}
 \caption{
Comparison between the data used by Lema\^{\i}tre in 1927 (left) to yield the
first empirical value of the rate of expansion of the Universe as 575 km/s/Mpc
(reconstructed in [30]), and the radial velocity--distance diagram published by
Hubble in 1929, with a best slope of 530 km/s/Mpc (right).
 }
 \end{center}
 \end{figure}

\item Section 6

The item 4 of the 1927's conclusions, giving the radius of the universe as 1/5th
the radius of Einstein's hypersphere, is suppressed, and in the next sentence,
Lema\^{\i}tre changes the range of the 100-inch Mount Wilson telescope estimated
by Hubble from $R/120$ to $R/200$.

\item Added references

Whereas the 1931 translation does not contain footnotes, it provides at the end
new references that could not be given in the 1927 article:  to Friedmann's article of 1922 and Einstein's comments on it, the article of Tolman
about models of variable radius of 1923, the developments of his own model given
by Eddington, De Sitter and himself in 1930, and eventually two popular
expositions given by him in 1929 (in French) and by De Sitter in 1931.

\end{itemize}

\bigskip

{\bf References}

\bigskip

\parindent=-0.5cm

\addtolength{\leftmargin}{0.5cm}

[1] J.-P. Luminet, \textit{Editorial note to  The beginning of the world from
the point of view of quantum theory}, \textit{Gen. Relativ. Gravit.} {\bf43},
n$^\circ$10, 2911-2928 (2011).

[2] H. Weyl,  \textit{Zur allgemeinen relativit\"{a}tstheorie},
\textit{Physikalische Zeitschrift}, {\bf 24}, 230--232 (1923). English
translation:  H. Weyl, \textit{On the general relativity theory}, \textit{Gen.
Relativ. Gravit.}, {\bf 35}, 1661 Ð 1666 (2009).

[3] C. Wirtz, \textit{De Sitters Kosmologie und die Radialbewegungen der
Spiralnebel}, \textit{Astronomische Nachtrichten}, {\bf 222}, 21 (1924).

[4] L. Silberstein, \textit{The Curvature of de Sitter's Space-Time Derived from
Globular Clusters}, \textit{M.N.R.A.S.}, {\bf 4} 363 (1924).

[5] K. Lundmark, \textit{The Motions and the Distances of Spiral Nebulae},
\textit{M.N.R.A.S.}, {\bf 85}, 865 (1925).

[6] H. S. Kragh, D. Lambert, \textit{The Context of Discovery: Lema\^{\i}tre and
the Origin of the Primeval-Atom Universe}, \textit{Annals of Science}, {\bf 64},
445--470 (2007).

[7] W. H. de Sitter, \textit{On Einstein's Theory of Gravitation and Its
Astronomical Consequences}, \textit{M.N.R.A.S.}, {\bf 78}, 3--28 (1917).

[8] G. Lema\^{\i}tre, \textit{Note on De Sitter's Universe}, \textit{The
Physical Review}, {\bf 25}, 903 (1925).

[9] H. Bondi, T. Gold, \textit{The Steady State Theory of the Expanding
Universe}, \textit{M.N.R.A.S.}, {\bf 108}, 252--270 (1948); F. Hoyle, \textit{A
New Model for the Expanding Universe}, \textit{M.N.R.A.S.}, {\bf 108}, 372--382
(1948).

[10] A. Linde, \textit{Particle Physics and Inflationary Cosmology}, Harwood,
1990.

[11] G. Lema\^{\i}tre, \textit{Note on De Sitter's Universe}, \textit{Journal of
Mathematics and Physics} {\bf 4} 189-192 (1925).

[12] G. Lema\^{\i}tre, \textit{The gravitational field in a fluid sphere of
uniform invariant density according to the theory of relativity; Note on de
Sitter Universe; Note on the theory of pulsating stars.}, Ph.D. thesis,
Department of Physics, M.I.T. (1927).

[13]  K. Freeman, \textit{Slipher and the Nature of the Nebulae}, in
\textit{Origins of the Expanding Universe: 1912-1932}, M. J. Way and D. Hunter,
eds., ASP Conf. Ser., Vol. 471 (2013), arXiv:1301:7509

[14] J. A. Peacock, \textit{Slipher, galaxies, and cosmological velocity
fields}, in \textit{Origins of the Expanding Universe: 1912-1932}, M. J. Way and
D. Hunter, eds., ASP Conf. Ser., Vol. 471 (2013), arXiv:1301.7286

[15] A . Eddington, \textit{The Mathematical Theory of Relativity}, Cambridge
University Press (1923), p.162 (reprinted 1960).

[16] M. Lachi\`eze-Rey,  J.-P. Luminet, \textit{Cosmic Topology},
\textit{Physics Reports}, {\bf 254}, 135--214 (1985).

[17] A. Friedmann, \textit{\"{U}ber die Kr\"{u}mmung des Raumes},
\textit{Zeitschrift f\"{u}r Physik}, {\bf 10}, 377 (1922). English translation:
\textit{On the curvature of space}, \textit{Gen. Relativ. Gravit.}, {\bf 31},
n$^\circ$12, 1991 (1999)

[18] A. Einstein, \textit{Notiz zu der Arbeit von A. Friedmann \"{U}ber die
Kr\"{u}mmung des Raumes}, \textit{Zeitschrift f\"{u}r Physik}, {\bf 16}, 228
(1923).

[19] O. Godart, \textit{Monseigneur Lema\^{\i}tre, sa vie, son oeuvre}, with
notes by Andr\'e Deprit, \textit{Revue des Questions Scientifiques}, {\bf  155},
155--182  (1984).

[20] M. Lachi\`eze-Rey,  J.-P. Luminet, \textit{Figures du Ciel}, Paris,
Seuil/BnF, 1998, p. 155. English translation: \textit{Celestial Treasury},
Cambridge University Press,  2001.

[21] G. Lema\^{\i}tre, \textit{La grandeur de l'espace}, \textit{Revue des
Questions Scientifiques}, {\bf 15}, 9--36 (1929).

[22]G. Str\"{o}mberg, \textit{Analysis of Radial Velocities of Globular Clusters
and Non-Galactic Nebulae}, \textit{Astrophysical Journal},  {\bf 61}, 353
(1925).

[23] E. Hubble,\textit{A Relation between Distance and Radial Velocity among
Extra-Galactic Nebulae}, Contributions from the Mount Wilson Observatory /
Carnegie Institution of Washington, {\bf 310} 1, (1927).

[24] J. Hopmann, \textit{Photometrische Untersuchungen von Nebelflecken},
\textit{Astronomische Nachrichten}, {\bf 214} 425, (1921).

[25] G. Shaviv \textit{Did Edwin Hubble plagiarize?}, ArXiv:1107.0442, (2011).

[26] D. Lambert, \textit{A propos de la controverse ``Hubble-- Lema\^{\i}tre''},
\textit{Pour la Science}, {\bf 412}, 78--81 (f\'evrier 2012).

[27] G. Lema\^{\i}tre, \textit{Rencontres avec Einstein}, \textit{Revue des
Questions Scientifiques} t. 79 5e s\'erie, vol. {\bf 19}, p. 129--132 (1958).

[28]  J.-P. Luminet, \textit{L'invention du big bang}, \textit{Le Seuil}, Paris,
2004; A. Belenkiy, \textit{Alexander Friedman and the Origins of Modern
Cosmology}, in \textit{Origins of the Expanding Universe: 1912-1932}, M. J. Way
and D. Hunter, eds., ASP Conf. Ser., Vol. 471 (2013), arXiv:1302.1498.

[29] H. P. Robertson, \textit{On Relativistic Cosmology}, \textit{Philosophical
Magazine} Series 7, { \bf 5}, 835 (1928).

[30] H. W. Duerbeck, W.C. Seitter, \textit{In HubbleÕs Shadow: Early Research on
the Expansion of the Universe},  Eds. C. Sterken and J.B. Hearnshaw, chap 15,
231 (2000). URL http://books.google.se/books?id=moPvAAAAMAAJ.

[31] H. P. Robertson, \textit{Proceedings of the National Academy of Sciences},
{\bf 15}, 822 (1929).

[32] E. Hubble,\textit{A Relation between Distance and Radial Velocitiy among
Extra-Galactic Nebulae}, \textit{Proceedings of the National Academy of
Sciences},  {\bf 15}, March 15, Number 3, 168 (1929).

[33] E. Hubble, \textit{The Realm of the Nebulae}, Yale University Press, 1936;
Dover Publications, 1958.

[34] G. E. Christianson, \textit{Edwin Hubble: Mariner of the Nebulae},
University of Chicago Press, 1996, p. 201.

[35] S. van den Bergh, \textit{Discovery of the Expansion of the Universe},
\textit{Journal of the Royal Astronomical Society of Canada}, {\bf 105}(5), 197 (2011) [arXiv:1108.0709v2].

[36] D. Block, \textit{A Hubble Eclipse: Lema\^{\i}tre and Censorship}, arXiv
:1106.3928 (2011).

[37]  R. C. Tolman, \textit{On the possible line elements for the universe}
\textit{Proceedings of the National Academy of Sciences}  {\bf 15}, 297 (1929).

[38] R. C. Tolman, \textit{Proceedings of the National Academy of Sciences}
{\bf 16}, 320 (1930).

[39] A.S. Eddington, \textit{Remarks at the Meeting of the Royal Astronomical
Society}, \textit{The Observatory}, vol. {\bf 53}, 39--40 (1930).

[40] H. Kragh, \textit{ Cosmology and Controversy}, Princeton University Press,
1996, chap.2.

[41] P.J.E. Peebles, \textit{Principes of Physical Cosmology}, Princeton
University Press, 1993, p. 80.

[42] J.-P. Luminet, \textit{Alexandre Friedmann, Georges Lema\^{\i}tre: Essais
de Cosmologie}, Paris, Le Seuil, coll. ``Sources du Savoir'', 1997.

[43] W. De Sitter, \textit{On the Distances and Radial Velocities of
Extra--galactic Nebulae, and the Explanation of the Latter by the Relativity
Theory of Inertia}, \textit{Proceedings of the National Academy of Sciences}
{\bf 16}(7), 474--488 (1930).

[44] A.S. Eddington, \textit{The Observatory}, vol. {\bf 53}, 162--164 (1930).

[45] A.S. Eddington, \textit{On the Instability of Einstein's Spherical World},
\textit{M.N.R.A.S.}  {\bf 90}, 668--678 (1930).

[46]  G. Lema\^{\i}tre, \textit{A homogeneous universe of constant mass and
increasing radius accounting for the radial velocity of extra-galactic nebulae},
\textit{M.N.R.A.S.}, {\bf 41}, 483--490 (1931).

[47] J.-P. Luminet, \textit{Editorial note to ``The beginning of the world from
the point of view of quantum theory''}, \textit{Gen. Relativ. Gravit.}, {\bf
43}, 2911-2928 (2011) [arXiv:1105.6271v1].

[48] S.L. Jaki, \textit{Science and Creation}, Scottish Academic Press,
Edinburgh and London, (1974); R. W. Smith, \textit{The expanding Universe,
Astronomy's ``Great Debate'' 1900-1931}, Cambridge, Cambridge University Press,
1982; O. Godart, M. Heller,  \textit{Cosmology of Lema\^{\i}tre}, Pachart Pub.
House, Tucson, 1985; J.P. Luminet, \textit{op. cit.},1997;  D. Lambert,
\textit{Un atome d'univers. La vie et l'\oe uvre de Georges Lema\^{\i}tre},
Bruxelles, Lessius, 2000; H. Kragh, R.W. Smith, \textit{Who discovered the
expanding Universe ?}, \textit{History of Science} {\bf 41}, 141 (2003); J.
Farrell, \textit{The Day Without Yesterday: Lema\^{\i}tre, Einstein, and the
Birth of Modern Cosmology},  New York, ThunderÕs Mouth Press, 2005; H.
Nussbaumer, L. Bieri,  \textit{Discovering the Expanding Universe}, Cambridge
University Press, 2009.

 [49] P. Couderc, \textit{L'expansion de l'Univers}, Paris, P.U.F., 1950.

[50] D. Lambert, \textit{Georges Lema\^{\i}tre: Rep\`eres biographiques},
\textit{Rev. Quest. Sci.} {\bf 183(4)}, 337--398 (2012).

[51] A. Kras\'inski, G.F.R. Ellis, \textit{Editor's note}, \textit{Gen. Relativ. Gravit.}, {\bf 31},
n$^\circ$12, 1985 (1999) 

[52] S. van den Bergh, \textit{The Curious Case of Lema\^{\i}tre's Equation
No.24}, \textit{Journal of the Royal Astronomical Society of Canada} {\bf 105}, 151 (2011)
[arXiv:1106.1195]; E. S. Reich, \textit{Edwin Hubble in translation trouble},
\textit{Nature} http://dx.doi.org/10.1038/news.2011.385 (2011).

[53] M. Livio, \textit{Mystery of the missing text solved}, \textit{Nature}, 10
November 2011, {\bf 479}, 171--173 (2011).

[54] G. Lema\^{\i}tre, \textit{The expanding universe}, \textit{M.N.R.A.S.},
{\bf 41}, 491--501 (1931).

[55] W.  de Sitter, \textit{On
the curvature of space}, \textit{Proceedings of the Royal Academy of Amsterdam},
{\bf 20}, 229 (1917).

[56] G. Lema\^{\i}tre, \textit{Compte-rendu du livre ``L'univers en expansion''
de Paul Couderc}, \textit{Annales d'Astrophysique}, {\bf XIII}(3), 344--345
(1950).

\newpage

\vskip2cm

\parindent=0.5cm

\begin{center}

{\bf APPENDIX}

\bigskip

{\bf A HOMOGENEOUS UNIVERSE OF CONSTANT MASS 
AND INCREASING RADIUS, 
ACCOUNTING FOR 
THE RADIAL VELOCITY OF EXTRAGALACTIC NEBULAE}

\vskip0.5cm

{\small Note by Abb\'e {\sc G. Lema\^{\i}tre}\\(Translation from the French original by J.-P. Luminet)}

\end{center}

{\bf {\sc 1. Generalities.}}

According to the theory of relativity, a homogeneous universe may exist not only
when the distribution of matter is uniform, but also when all positions in space
are completely equivalent, there is no center of gravity. The radius $R$ of space
is constant, space is elliptic with uniform positive curvature $1/R^2$, the lines
starting from a same point come back to their starting point after having
travelled a path equal to $\pi R$, the total volume of space is finite and equal
to $\pi^2 R^3$, straight lines are closed lines going through the whole
 space without encountering any boundary($^1$).\footnotetext[1]{We consider simply connected elliptic space,
i.e. without antipodes.}

Two solutions have been proposed. That of {\sc de Sitter} ignores the existence
of matter and supposes its density equal to zero. It leads to special
difficulties of interpretation which we will be referred to later, but it is of
great interest as explaining the fact that extragalactic nebul\ae\  seem to
recede from us with a huge velocity, as a simple consequence of the properties
of the gravitational field, without having to suppose that we are at a point of
the universe distinguished by special properties.

The other solution is that of {\sc Einstein}. It pays attention to the obvious
fact that the density of matter is not zero and it leads to a relation between
this density and the radius of the universe. This relation forecasted the
existence of masses enormously greater than any known when the theory was for
the first time compared with the facts. These masses have since been discovered,
the distances and dimensions of extragalactic nebul\ae\  having become
established. From Einstein's formula and recent observational data, the radius
of the universe is found to be some hundred times greater than the most distant
objects which can be photographed by our telescopes ($^2$).\footnotetext[2]{
Cf. Hubble E. Extra-Galactic Nebul\ae, {\it ApJ.}, vol. 64, p. 321, 1926. {\it
M$^t$ Wilson Contr.} N$^{\circ}$ 324.}

Each theory has its own advantage. One is in agreement with the observed radial
velocities of nebul\ae, the other with the existence of matter, giving a
satisfactory relation between the radius and the mass of the universe. It seems
desirable to find an intermediate solution which could combine the advantages of
both.

At first sight, such an intermediate solution does not appear to exist. A static
gravitational field with spherical symmetry has only two solutions, that of
Einstein and that of de Sitter, if the matter is uniformly distributed without
pressure  or internal stress. De Sitter's universe is empty, that of Einstein
has been described as containing as much matter as it can contain. It is
remarkable that the theory can provide no mean between these two extremes.

The solution of the paradox is that the de Sitter's solution does not really meet
all the requirements of the problem ($^3$).\footnotetext[3]{Cf. {\sc K.
Lanczos}, Bemerkung zur de Sitterschen Welt, {\it Phys. Zeitschr.} vol. 23, 1922,
p. 539, and {\sc H. Weyl}, Zur allgemeinen Relativit\"{a}tstheorie, Id., vol. 24,
1923, p. 230, 1923. We follow the point of view of Lanczos here. The worldlines
of nebul\ae\ form a bunch with ideal center and real axial hyperplane; space
orthogonal to these worldlines is formed by the hyperspheres equidistant from the
axial plane. This space is elliptic, its variable radius being minimum at the
moment corresponding to the axial plane. Following the assumption of Weyl, the
worldlines are parallel in the past; the normal hypersurfaces representing space
are horospheres, the geometry of space is thus Euclidean. The spatial distance
between nebul\ae\  increases as the parallel geodesics which they follow
 recede one from  the other proportionally to $e^{t/R}$, where $t$ is the proper time and $R$
the radius of the universe. The Doppler effect is equal to $r/R$, where $r$ is
the distance from the source at the moment of observation. Cf. {\sc G.
Lema\^{\i}tre}, Note on de Sitter's universe, {\it Journal of mathematics and
physics}, vol. 4, n$^{\circ}$3, May 1925, or {\it Publications du Laboratoire
d'Astronomie et de G\'eodesie de l'Universit\'e de Louvain}, vol. 2, p. 37, 1925.
For the discussion of the de Sitter's partition, see {\sc P. Du Val}, Geometrical
note on de Sitter's world, {\it Phil. Mag.} (6), vol. 47, p. 930, 1924. Space is
constituted by hyperplanes orthogonal to a time line described by the introduced
center, the trajectories of nebul\ae\ are the trajectories orthogonal to these
planes, in general they are no more geodesics and they tend to becoming lines of
null length when one approaches the horizon of the center, i.e. the polar
hyperplane of the central axis with respect to the absolute one.} Space is
homogeneous with constant positive curvature; space-time is also homogeneous, for
all events are perfectly equivalent. But the partition of space-time into space
and time disturbs the homogeneity. The selected coordinates introduce a center to
which nothing corresponds in reality; a particle at rest somewhere else
  than at the center does not describe a geodesic. The coordinates chosen destroy
the homogeneity that exists in the data for the problem and produce the paradoxical results which appear at the so-called "<horizon">
of the center. When we use coordinates and a corresponding partition of space
and time of such a kind as to preserve the homogeneity of the universe, the
field is found to be no longer static ; the universe becomes of the same form as
that of Einstein, with a radius of space no longer constant but varying with the
time according to a particular law ($^4$).\footnotetext[4]{If we restrict
the problem to two dimensions, one of space and one of time, the partition of
space and time used by Sitter can be represented on a sphere: the lines of space
are provided by a system of great circles which intersect on a same diameter,
and the lines of time are the parallels cutting
orthogonally the lines of space. One of these parallels is a great circle and thus a
geodesic, it corresponds to the center of space, the pole of this great circle is
a singular point corresponding to the horizon of the center. Of course the
representation must be extended to four dimensions and the time coordinate must
be assumed imaginary, but the defect of homogeneity resulting from the choice of
the coordinates remains. The coordinates respecting the homogeneity require
 taking a system of meridian lines as lines of time and the corresponding
parallels for lines of space, whereas the radius of space varies with time.}

In order to find a solution combining the advantages of those of Einstein and de
Sitter, we are led to consider an Einstein universe where the radius of space
(or of the universe) is allowed to vary in an arbitrary way.

\bigskip

{\bf \sc 2. Einstein universe of variable radius. Field equations. Conservation
of energy.}

As in Einstein's solution, we liken the universe to a rarefied gas whose
molecules are the extragalactic nebul\ae\ . We suppose them so numerous that a
volume small in comparison with the universe as a whole contains enough
nebul\ae\ to allow us to speak of the density of matter. We ignore the possible
influence of local condensations. Furthermore, we suppose that the nebul\ae\ are
uniformly distributed so that the density does not depend on position.

When the radius of the universe varies in an arbitrary way, the density, uniform
in space, varies with time. Furthermore, there are generally internal stresses
which, in order to preserve the homogeneity, must reduce to a simple pressure,
uniform in space and variable with time. The pressure, being two-thirds of the
kinetic energy of the molecules, is negligible with respect to the energy
associated with matter; the same can be said of interior stresses in nebul\ae\ or
in stars belonging to them. We are thus led to put $p = 0$. Nevertheless it might
be necessary to take into account the radiation-pressure of electromagnetic
energy travelling through space; this energy is weak but it is evenly distributed
through the whole of space and might provide a notable contribution to the mean energy. We shall keep the pressure $p$ in the
general equations as the average radiation-pressure of light, but we shall write
$p = 0$ when we discuss the application to astronomy.

We denote the density of total energy by $\rho$, the density of radiation energy
by $3p$, and the density of the energy condensed in matter by $\delta = \rho -
3p$.

We identify $\rho$ and $- p$ with the components $T_4^4$ and $T^1_1 = T^2_2 =
T^3_3$ of the material energy tensor, and $\delta$ with $T$. Working out the
contracted Riemann tensor for a universe with a line-element given by
 $$
ds^2 = - R^2 d \sigma^2 + dt^2 \eqno{(1)}
 $$
where $d \sigma$ is the elementary distance in a space of radius unity, and
 the radius of space $R$ is a function of time,
 we find that the field equations can be written
 $$
3 \frac {{R'}^2} {R^2} + \frac 3 {R^2} = \lambda + \kappa \rho \eqno{(2)}
 $$
and
 $$
2 \frac {R''} R + \frac {{R'}^2} {R^2} + \frac 1 {R^2} = \lambda - \kappa \rho
\eqno{(3)}
 $$

Accents denote derivatives with respect to $t$; $\lambda$ is
the cosmological constant whose value is unknown,
 and $\kappa$ is the Einstein constant whose value is $1,87 \times 10^{-27}$ in
C.G.S. units ($8 \pi$ in natural units).

The four identities expressing the conservation of momentum and of energy reduce
to
 $$
\frac {d \rho} {dt} + \frac {3R'} R (\rho + p) = 0 \eqno(4)
 $$
which is the conservation of energy equation. This equation can replace (3). It
is suitable for an interesting interpretation. Introducing the volume of space
$V = \pi^2 R^3$, it can be written
 $$
d(V \rho) + p dV = 0  \eqno(5)
 $$
showing that {\em the variation of total energy plus the work done by
radiation-pressure is equal to zero}.

\bigskip

{\bf \sc 3. Case of a universe of constant total \sc mass.}

Let us seek a solution for which the total mass $M = V \delta$ remains constant.
We can write
 $$
\kappa \delta = \frac {\alpha} {R^3} \eqno{(6)}
 $$
where $\alpha$ is a constant. Taking account of the relation
 $$
\rho = \delta + 3 p
 $$
existing between the various kinds of energy, the principle of conservation of
energy becomes
 $$
3 d(p R^3) + 3 p R^2 dR = 0 \eqno{(7)}
 $$
whose integration is immediate; and, $\beta$ being a constant of integration,
 $$
\kappa p = \frac {\beta} {R^4} \eqno{(8)}
 $$
and therefore
 $$
\kappa \rho = \frac {\alpha} {R^3} + \frac {3 \beta} {R^4} \eqno{(9)}
 $$
By substitution in (2) we have to integrate
 $$
\frac {{R'}^2} {R^2} = \frac {\lambda} 3 - \frac 1 {R^2} + \frac {\kappa \rho} 3
= \frac {\lambda} 3 - \frac 1 {R^2} + \frac {\alpha} {3 R^3} + \frac {\beta}
{R^4} \eqno{(10)}
 $$
or
 $$
t = \int \frac {dR} {\sqrt{\frac {\lambda R^2} 3 - 1 + \frac {\alpha} {3 R} +
\frac {\beta} {R^2}}} \eqno{(11)}
 $$

When $\alpha$ and $\beta$ vanish, we obtain the de Sitter solution
($^5$)\footnotetext[5]{Cf. {\sc Lanczos}, {\it l.c.}}
 $$
R = \sqrt{\frac 3 {\lambda}}\ \cosh \sqrt{\frac {\lambda} 3}\ (t - t_0)
\eqno{(12)}
 $$

The Einstein solution is found by making $\beta = 0$ and $R$ constant. Writing
$R' = R'' = 0$ in (2) and (3) we find
 $$
\frac 1 {R^2} = \lambda \qquad \frac 3 {R^2} = \lambda + \kappa \rho \qquad \rho
= \delta
 $$
or
 $$
R = \frac 1 {\sqrt{\lambda}} \qquad \kappa \rho = \frac 2 {R^2} \eqno{(13)}
 $$
and from (6)
 $$
\alpha = \kappa \delta R^3 = \frac 2 {\sqrt{\lambda}} \eqno{(14)}
 $$

The Einstein solution does not result from (14) alone, it also supposes that the
initial value of $R'$ is zero. Indeed, if, in order to simplify the notation,
we write
 $$
\lambda = \frac 1 {R_0^2} \eqno{(15)}
 $$
and put in (11) $\beta = 0$ and $\alpha = 2R_0$, it follows that
 $$
t = R_0 \sqrt{3} \int \frac {dR} {R - R_0} \sqrt{\frac R {R + 2R_0}} \eqno{(16)}
 $$

For this solution the two equations (13) are of course no longer valid. Writing
 $$
\kappa \delta = \frac 2 {R_E^2} \eqno{(17)}
 $$
we have from (14) and (15)
 $$
R^3 = R_E^2 R_0 \eqno{(18)}
 $$

The value of $R_E$, the radius of the universe computed from the average density
by Einstein's equations (17), has been found by Hubble to be
 $$
R_E = 8,5 \times 10^{28} {\rm cm.} = 2,7 \times 10^{10} {\rm parsecs}
\eqno{(19)}
 $$

We shall see later that the value of $R_0$ can be computed from the radial
velocities of the nebul\ae; $R$ can then be found from (18). Finally, we shall
show that a solution introducing a relation substantially different
 from (14) would lead to consequences not easily acceptable.

\bigskip

{\bf \sc 4. Doppler effect due to the variation of the radius of the universe}

{}From (1) giving the line element of the universe,
 the equation for a light ray is
 $$
\sigma_2 - \sigma_1 = \int_{t_1}^{t_2} \frac {dt} R \eqno{(20)}
 $$
where $\sigma_1$ and $\sigma_2$ relate to spatial coordinates.
 We suppose that the light is emitted at the point $\sigma_1$ and observed at
$\sigma_2$.
 A ray of light emitted slightly later starts from $\sigma_1$ at time $t_1 +
\delta t_1$ and reaches $\sigma_2$ at time $t_2 + \delta t_2$. We have therefore
 $$
\frac {\delta t_2} {R_2} - \frac {\delta t_1} {R_1} = 0, \qquad \frac {\delta
t_2} {\delta t_1} - 1 = \frac {R_2} {R_1} - 1 \eqno{(21)}
 $$
 where $R_1$ and $R_2$ are the values of the radius $R$ at the time of emission
$t_1$ and at the time of observation $t_2$.
$t$ is the proper time; if $\delta t_1$ is the period of the emitted light,
$t_2$ is the period of the observed light.
Moreover, $\delta t_1$ can also be considered as
 the period of the light emitted under the same conditions in the neighbourhood
of the observer, because the period of the light emitted under the same
physical conditions has the same value everywhere when reckoned in proper time.
Therefore
 $$
\frac v c = \frac {\delta t_2} {\delta t_1} - 1 = \frac {R_2} {R_1} - 1
\eqno{(22)}
 $$
measures the apparent Doppler effect due to the variation of the radius of the
universe. {\em It equals the ratio of the radii of the universe at the instants
of observation and emission, diminished by unity.} $v$ is that velocity of the
observer which would produce the same effect. When the source is near enough, we
can write approximately
 $$
\frac v c = \frac {R_2 - R_1} {R_1} = \frac {dR} R = \frac {R'} R dt = \frac
{R'} R r
 $$
where $r$ is the distance of the source. We have therefore
 $$
\frac {R'} R = \frac v {cr} \eqno{(23)}
 $$

Radial velocities of 43 extragalactic nebul\ae\  are given by Str\"{o}mberg
($^6$).\footnotetext[6]{Analysis of radial velocities of globular clusters
and non galactic nebul\ae. {\it Ap.J.} vol. 61, p. 353, 1925. {\it M$^t$ Wilson
Contr.}, N$^{\circ}$ 292.}

The apparent magnitude $m$ of these nebul\ae\  can be found in the work of
Hubble. It is possible to deduce their distance from it, because Hubble has shown
that extragalactic nebul\ae\  have approximately equal absolute magnitudes
(magnitude $= - 15.2$ at 10 parsecs, with individual variations $\pm 2$), the
distance $r$ expressed in parsecs is then given by the formula $\log r = 0,2 m +
4,04$.

One finds a distance of about $10^6$ parsecs, varying from a few tenths to 3,3
megaparsecs. The probable error resulting from the dispersion of absolute
magnitudes is considerable. For a difference in absolute magnitude of $\pm 2$,
the distance exceeds from 0,4 to 2,5 times the calculated distance. Moreover,
the error is proportional to the distance. One can admit that, for a distance of
one megaparsec, the error resulting from the dispersion of magnitudes is of the
same order as that resulting from the dispersion of velocities. Indeed, a
difference of magnitude of value unity corresponds to a proper velocity of 300
Km/s, equal to the proper velocity of the sun compared to nebul\ae\ . One can
hope to avoid a systematic error by giving to the observations a weight
proportional to $\frac 1 {\sqrt{1 + r^2}}$, where $r$ is the distance in
megaparsecs.

Using the 42 nebul\ae\  appearing in the lists of Hubble and Str\"{o}mberg
($^7$),\footnotetext[7]{Account is not taken
 of N.G.C. 5194 which is associated with N.G.C. 5195. The introduction of the
Magellanic clouds would be without influence on the result.} and taking account
of the proper velocity of the Sun (300 Km/s in the direction $\alpha =
315^{\circ}$, $\delta = 62^{\circ}$), one finds a mean distance of 0,95
megaparsecs and a radial velocity of 600 Km/sec, i.e. 625 Km/sec at $10^6$
parsecs ($^8$).\footnotetext[8]{By not giving a weight to the observations,
one would find 670 Km/sec at $1.16 \times 10^6$ parsecs, 575 Km/sec at $10^6$
parsecs. Some authors sought to highlight the relation between $v$ and $r$ and
obtained only a very weak correlation between these two terms. The error in the
determination of the individual distances is of the same order of magnitude as
the interval covered by the observations and the proper velocity of nebul\ae\
(in any direction) is large (300 Km/sec according to Str\"{o}mberg), it thus
seems that these negative results are neither for nor against the relativistic
interpretation of the Doppler effect. The inaccuracy of the observations makes
only possible to assume $v$ proportional to $r$ and to try to avoid a systematic
error in the determination of the ratio $v/r$. Cf. {\sc Lundmark}, The
determination of the curvature of space time in de Sitter's world, M.N., vol.
84, p. 747, 1924, and {\sc Str\"{o}mberg}, {\it l.c.} }

We will thus adopt
 $$
\frac {R'} R = \frac v {rc} = \frac {625 \times 10^5} {10^6 \times 3,08 \times
10^{18} \times 3 \times 10^{10}} = 0,68 \times 10^{-27} {\rm cm}^{-1}
\eqno{(24)}
 $$

This relation enables us to calculate $R_0$. We have indeed by (16)
 $$
\frac {R'} R = \frac 1 {R_0 \sqrt{3}} \sqrt{1 - 3 y^2 + 2 y^3} \eqno{(25)}
 $$
where we have set
 $$
y = \frac {R_0} R \eqno{(26)}
 $$
On the other hand, from (18) and (26)
 $$
R_0^2 = R_E^2 y^3 \eqno{(27)}
 $$
and therefore
 $$
3 \left(\frac {R'} R\right)^2 R_E^2 = \frac {1 - 3y^2 + 2y^3} {y^3} \eqno{(28)}
 $$

With the adopted numerical data (24) for $\frac {R'} R$ and (19) for $R_E$, we
have
 $$
y = 0,0465.
 $$
We have therefore
 $$
R = R_E \sqrt{y} = 0,215 R_E = 1,83 \times 10^{28} {\rm cm.} = 6 \times 10^9
{\rm \ parsecs}
 $$
 $$
R_0 = Ry = R_E y^{\frac 3 2} = 8,5 \times 10^{26} {\rm cm.} = 2,7 \times 10^8
{\rm \ parsecs}
 $$
 $$
= 9 \times 10^8 {\rm \ light\ years.}
 $$

Integral (16) can easily be computed. Writing
 $$
x^2 = \frac R {R + 2R_0} \eqno{(29)}
$$
it can be written
 $$
t = R_0 \sqrt{3} \int \frac {4 x^2 dx} {(1 - x^2)(3x^2 - 1)} = R_0 \sqrt{3} \log
\frac {1+ x} {1 - x} + R_0 \log \frac {\sqrt{3} x - 1} {\sqrt{3} x + 1} + C
\eqno{(30)}
 $$

If $\sigma$ is the fraction of the radius of the universe travelled by light
during time $t$, we have also from (20)
 $$
\sigma = \int \frac {dt} R = \sqrt{3} \int \frac {2 dx} {3x^2 - 1} = \log \frac
{\sqrt{3} x - 1} {\sqrt{3} x + 1} + C'. \eqno{(31)}
 $$

The following table gives values of $\sigma$ and $t$ for different values of
$\frac R {R_0}$.

\begin{center}
\begin{tabular}{||r|c|c|c||}
 \hline \hline
 $\frac R {R_0}$ & $\frac t {R_0}$ &
 \hspace{-5.75mm}
  \begin{tabular}{c}
      $\sigma$ \\
      \begin{tabular}{c|c}
      \hline
      {\small RADIANS} & {\small DEGREES} \\
      \end{tabular}
  \end{tabular}
      \hspace{-5.5mm} & $\frac v c$ \\
 \hline \hline
 \  & \  & \hspace{-2.75mm}
    \begin{tabular}{c|c}
     \  & \  \\
     \end{tabular} & \  \\
 1  &  \hspace{-4mm} $-\infty$  & \hspace{-1.25mm}
    \begin{tabular}{l|l}
    \hspace{-10mm} $-\infty$ \hspace{7.75mm} & \hspace{0.5mm} $-\infty$ \\
    \end{tabular} &  19 \\
 2  & \ $- 4,31$  & \hspace{-1.25mm}
    \begin{tabular}{l|l}
    \hspace{-8mm} $- 0,889$ \hspace{1.5mm} & \hspace{0.5mm} $- 51^{\circ}$ \\
    \end{tabular} &  9 \\
 3  & \ $- 3,42$  & \hspace{-1.25mm}
    \begin{tabular}{l|l}
    \hspace{-8mm} $- 0,521$ \hspace{1.5mm} & \hspace{0.5mm} $- 30^{\circ}$ \\
    \end{tabular} & \hspace{2mm} $5 \frac 2 3$ \\
 4  & \ $- 2,86$  & \hspace{-1.25mm}
    \begin{tabular}{l|l}
    \hspace{-8mm} $- 0,359$ \hspace{1.5mm} & \hspace{0.5mm} $- 21^{\circ}$ \\
    \end{tabular} & \  4 \\
 5  & \ $- 2,45$  & \hspace{-1.25mm}
    \begin{tabular}{l|l}
    \hspace{-8mm} $- 0,266$ \hspace{1.5mm} & \hspace{0.5mm} $- 15^{\circ}$ \\
    \end{tabular} & \  3 \\
 10 & \ $- 1,21$  & \hspace{-1.25mm}
    \begin{tabular}{l|l}
    \hspace{-7.5mm} $- 0,087$ \hspace{1mm} & \hspace{0.5mm} $- \hspace{2mm} 5^{\circ}$ \\
    \end{tabular} & \  1 \\
 15 & \ $- 0,50$  & \hspace{-1.25mm}
    \begin{tabular}{l|l}
    \hspace{-5.5mm} $- 0,029$ \hspace{1.5mm} & \hspace{0.5mm} $- \hspace{2.5mm} 1^{\circ}7$ \\
    \end{tabular} & \  $\frac 1 3$ \\
 20 & \hspace{-2.5mm} $0$  & \hspace{-1.25mm}
    \begin{tabular}{l|l}
    \hspace{-5.5mm} $0$ \hspace{9.25mm} & \hspace{6.5mm} $0$ \\
    \end{tabular} & \  0 \\
 25 & \hspace{3.5mm} $0,39$  & \hspace{-1.25mm}
    \begin{tabular}{l|l}
    \hspace{-4mm} $0,017$ \hspace{1mm} & \hspace{0.5mm} $ \hspace{5.5mm} 1^{\circ}$ \\
    \end{tabular} & \  \\
 $\infty$ & \hspace{0.0mm} $\infty$  & \hspace{-1.25mm}
    \begin{tabular}{l|l}
    \hspace{-4mm} $0,087$ \hspace{1mm} & \hspace{0.5mm} $ \hspace{5.5mm} 5^{\circ}$ \\
    \end{tabular} & \  \\
 \hline \hline
\end{tabular}
\end{center}

The constants of integration are adjusted to make $\sigma$ and $t$ vanish for
$\frac R {R_0} = 20$ in place of 21,5. The last column gives the Doppler effect
computed from (22). The approximate formula (23) would make $\frac v c$
proportional to $r$ and thus to $\sigma$. The error
committed by adopting this equation is only 0.005 for $\frac v c = 1$. The approximate formula 
may therefore be used within the limits of the visible spectrum.

\bigskip

{\bf \sc 5. The meaning of equation (14).}

The relation (14) between the two constants $\lambda$ and $\sigma$ has been
adopted following Einstein's solution. It is the necessary condition that quartic
under the radical in (11) may have a double root $R_0$ giving on integration a
logarithmic term. For simple roots, integration would give a square root,
corresponding to a minimum of $R$ as in de Sitter's solution (12). This minimum
would generally occur at time of the order of $R_0$, say $10^9$ years, i.e. quite
recently for stellar evolution. It thus seems that the relation existing between
the constants $\alpha$ and $\lambda$ must be close to (14) for which this minimum
is removed to the epoch at minus infinity ($^9$).
  \footnotetext[9]{If the positive roots were to become imaginary,
the radius would vary from zero upwards, the variation slowing down in the
neighbourhood of the modulus of the imaginary roots.
For a relation substantially different from (14), this slowing down becomes weak
and the time of evolution after leaving $R = 0$ becomes again of the order of
$R_0$.
 }

\bigskip

{\bf \sc 6. Conclusion.}

We have found a solution such that:

1. The mass of the universe is a constant related to the cosmological constant
by Einstein's relation
 $$
\sqrt{\lambda} = \frac {2 \pi^2} {\kappa M} = \frac 1 {R_0}
 $$

2. The radius of the universe increases without limits from an asymptotic value
$R_0$ for $t = - \infty$.

3. The recession velocities of extragalactic nebul\ae\ are a cosmical effect of the expansion of
the universe. The initial radius $R_0$ can be computed by formul\ae\ (24) and
(25) or by the approximate formula $R_0 = \frac {rc} {v \sqrt{3}}$.

4. The radius of the universe is of the same order of magnitude as the radius
$R_E$ deduced from density according to Einstein's formula
 $$
R = R_E \sqrt[3]{\frac {R_0} {R_E}} = \frac 1 5 R_E
 $$

This solution combines the advantages of the Einstein and de Sitter solutions.

Note that the largest part of the universe is forever out of our reach. The
range of the 100-inch Mount Wilson telescope is estimated by Hubble to be $5
\times 10^7$ parsecs, or about $\frac 1 {120} R$. The corresponding Doppler
effect is 3000 Km/sec. For a distance of $0,087 R$ it is equal to unity, and the
whole visible spectrum is displaced into the infra-red. It is impossible
that ghost images of nebul\ae\ or suns would form,
as even if there were no absorption these images would be displaced by several
octaves into the infra-red and would not be observed.

It remains to find the cause of the expansion of the universe. We have seen that
the pressure of radiation does work during the expansion. This seems to suggest
that the expansion has been set up by the radiation itself. In a static
universe, light emitted by matter travels round space, comes back to its
starting point and accumulates indefinitely. It seems that this may be the
origin of the velocity of expansion $R'/R$ which Einstein assumed to be zero and
which in our interpretation is observed as the radial velocity of extragalactic
nebul\ae.

\end{document}